\documentstyle[preprint,pra,aps]{revtex}        
\input epsf
 \begin{document}
 \draft

\title{New approach to the correlation spectrum near intermittency: 
a quantum mechanical analogy}
\author{Julius Bene$^{\scriptstyle (1)}$, Zolt\'an Kaufmann$^{\scriptstyle (1)}$ 
and Hans Lustfeld$^{\scriptstyle (2)}$}
\address{$^{\scriptstyle (1)}$Institute for Solid State Physics, E\"otv\"os University,
M\'uzeum krt. 6-8, H-1088 Budapest, Hungary\\ 
$^{\scriptstyle (2)}$Institut f\"ur Festk\"orperforschung, 
Forschungszentrum J\"ulich, D 52425 J\"ulich, Germany}

\date{\today}
\maketitle

\begin{abstract}
The correlation spectrum of fully developed one-dimensional mappings 
are studied near and at a weakly intermittent situation. Using a suitable 
infinite matrix representation, the eigenvalue equation of 
the Frobenius-Perron operator is approximately reduced to the 
radial Schr\"odinger equation of the hydrogen atom. 
Corrections are calculated by quantum mechanical perturbation
theory. Analytical expressions for the spectral properties
and correlation functions are derived and checked numerically.
Compared to our previous works, the accuracy of the present results
is significantly higher owing to the controlled and
systematic approximation scheme.
\end{abstract}
\pacs{PACS numbers: 05.45. +b}

\section{ Introduction}
Typical intermittent behavior involves an irregular 
switching between 
regular and chaotic motion. As a result, correlation decay in 
intermittent systems is intermediate between that of regular and `purely'
chaotic systems: it follows typically a power law\cite{GrossHorn}.
Hereafter we consider intermittency in case of
one-dimensional noninvertible mappings.  
Correlation functions may then be expressed in terms of the spectral properties of the
Frobenius-Perron operator $\hat H$, defined by
\begin{eqnarray}
\left(\hat H \varphi\right)(x)=\int \;dy\;\delta\left(x-f(y)\right)\varphi(y)\label{e1.0}\\
=\sum_a \frac{\varphi\left(f^{-1}_a(x)\right)}{\left|f'\left(f^{-1}_a(x)\right)\right|}\;.\nonumber
\end{eqnarray}
Here $ \varphi(x)$ is an arbitrary function, $f(x)$ stands for the mapping function
and $a$ labels the inverse branches.
The power law decay 
is associated with an accumulation point at the upper edge of the spectrum 
that causes serious difficulties at direct numerical studies of the spectral 
properties. Moreover, we have found that the corresponding 
eigenvectors tend to become approximately parallel, that greatly
reduces the number of eigenstates which can be extracted from
a finite size matrix representation\cite{KLB}. 
Therefore, a suitable analytical
treatment is needed which does not involve any truncations. Such a method 
has already been applied in Refs.\cite{KLB}, \cite{LBK}. Here we present another
approach which is based on a suitable infinite matrix representation.
This method has the advantage that it approximately reduces the problem 
to the well known radial Schr\"odinger equation for {\em s}-states.
Mathematically significant is the fact that the
original problem is transformed into an equivalent one which
admits a nearly 
hermitian representation. Therefore,
beyond the interesting analogy, our method renders possible the application
of standard perturbation theory as a tool for a systematic calculation
of corrections. This results in a rather high accuracy
for the spectral properties, as we shall demonstrate.

We develop our method for the treatment of the family of piecewise parabolic
maps\cite{ppmap}
\begin{eqnarray}
x'=\frac{1}{2r}\left(1+r-\sqrt{(1-r)^2+4r|1-2x|}\right)\quad r\in [0,1]\;.
\label{e1.1}
\end{eqnarray}
Note that the parameter value $r=1$ corresponds to a weakly intermittent 
situation \cite{weakint} - \cite{weakint4}, \cite{GrossHorn}. 
It turns out that 
the main features of the spectrum and the eigenfunctions depend only on 
the behavior of the map near the unstable fixed point $0$ when $r$ is 
close to $1$, thus our results have a broader relevance, in fact, they bear in a
sense a universal character\cite{LBK}.

The paper is organized as follows. In Section II the intermittent situation
($r=1$) is discussed, especially, contact is made with the quantum mechanical $s$-scattering
on a Coulomb potential at zero energy. 
In Section III the nearly intermittent ($r\approx 1$) situation is 
considered. The problem is reduced approximately to the bound $s$-states 
in a Coulomb potential. Correction terms are calculated by 
standard quantum mechanical perturbation theory.
Approaching intermittency, the number of the eigenstates essentially
involved in
the representation of a correlation function tends to the infinity,
therefore the calculation of correlation functions is highly nontrivial,
even if the spectral properties are already known\cite{LBK}.
In Section IV we present and discuss an analytic expression
for correlation functions, that is relevant 
near intermittency. The detailed derivation can be found 
in Appendix B.
A discussion of the results
is given in the concluding Section V. Appendix A contains 
an outline and discussion of the numerical method for calculating corrections 
in the intermittent situation.

\section{ The intermittent situation}
One conclusion of our numerical study done on the family of maps (\ref{e1.1})
 has been 
that finite size matrix approximations are poor near and at the intermittent 
situation, hence if one wants to get the spectral properties in those 
cases reliably, some other method is needed which does not involve any 
truncations of the matrix representations. As a first step we have 
chosen for an analytical study the intermittent map 
\begin{eqnarray}
x'=1-\sqrt{|1-2x|}\quad,\label{e2.1}
\end{eqnarray}
corresponding to $r=1$ in Eq.\ (\ref{e1.1}).
Using the basis $(1-x)^{4n+1}$ (where $n=0,1,2,...$), one obtains 
for the $(j,k)$-th matrix element of the Frobenius-Perron-operator (\ref{e1.0})
the simple closed expression
\begin{eqnarray} H_{j,k}=\left(
\begin{array}{c}
4k+1  \\
2j
\end{array} \right)2^{-4k}\quad.\label{e2.2}
\end{eqnarray}
These matrix elements are displayed in Fig. 1.
Our aim is to find 
an asymptotical solution to the eigenvalue equation, i.e. an expression 
of the eigenvector for large $j$-values, as the failure of the numerical 
calculation implies that this numerically inaccessible 
part of the eigenvector plays an inevitable role. The matrix 
representation (\ref{e2.2}) also supports this expectation as the largest matrix 
elements lie at the diagonal and decay with $j$ as $j^{-\frac{1}{2}}$, 
while the strip of the non-negligible matrix elements along the diagonal 
has a width of $\approx \sqrt{j}$. Thus any truncation leads to a 
dramatic effect. 
Using Stirling's
formula for the factorials arising in Eq.\ (\ref{e2.2})
for large $j$ and $|k-j|\approx \sqrt{j}$ the $H_{j,k}$ matrix
element is approximately given by
\begin{eqnarray}
H_{j,k}=
\sqrt{\frac{2}{\pi j} }\exp \left( -\frac{2m^2}{j}\right)
\left( 1+ \left( -\frac{3}{2}\frac{m}{j}+2\frac{m^3}{j^2}\right)
+\left( -\frac{5}{16}\frac{1}{j} +\frac{27}{8}\frac{m^2}{j^2}
-\frac{16}{3}\frac{m^4}{j^3} + 2\frac{m^6}{j^4} \right) \right)\label{e2.3}
\end{eqnarray}
 (where $m=k-j$), this
expression being accurate up to the order $\frac{1}{j}$. 
The quality of the approximation is demonstrated in Fig. 2, where
the difference between the asymptotical expression (\ref{e2.3}) and
the exact values (\ref{e2.2}) is shown.
Retaining only the Gaussian, we find the approximate eigenvectors
\begin{eqnarray}
a^{(1)}_k=\sin\left( C\sqrt{k}\right) \quad,\label{e2.4}
\end{eqnarray}
and
\begin{eqnarray}
a^{(2)}_k=\cos\left( C\sqrt{k}\right) \quad,\label{e2.5}
\end{eqnarray}
both belonging to the
eigenvalue
\begin{eqnarray}
\lambda=\exp\left(-\frac{C^2}{32}\right)\quad.\label{e2.6}
\end{eqnarray}
This can be proven by
inserting the above asymptotic expressions into the eigenvalue
equation, replacing the summation over $k$ with integration and
evaluating it by using the saddle point method. The reason why
we get this twofold degeneracy can be understood as a result of
the asymptotical method which formally allows $j$ and $k$ to
take on negative values as well, thus the situation is analogous
with the continuous spectrum of a quantum particle performing
unbounded one-dimensional motion. Actually these solutions get
corrections near the origin (i.e., at $k=0$), where the
asymptotic expansion of $H_{j,k}$ does not hold, and, on the
other hand they combine in a particular way to cancel each other
for $k<0$. As a result, the spectrum will not be degenerated,
but is continuous. Using a quantum mechanical analogy again, the
situation resembles to a one dimensional scattering process of a
particle on a potential containing a hard core.
It is interesting to note, that the asymptotic regime
corresponds to the laminar motion in the original $x$-representation, as for
large $k$ the basis functions $(1-x)^{4k+1}$ are sharply
peaked at the origin (which is now a marginally unstable fixed
point) and nearly vanish elsewhere. The nonasymptotic regime, on
the other hand, corresponds to the chaotic motion. Hence in this
representation we are given a picture about intermittency that
relates it to a scattering process, where the motion near the
scattering potential corresponds to chaos. As we shall see later,
in the nonintermittent situation
one has to do with bound states in such a potential. This is in accordance with 
the fact that we
have a discrete spectrum in that case.
 
Nevertheless, the above asymptotic solutions are not very
precise numerically. To improve this approximation, one may seek
for corrections proportional to $\frac{1}{\sqrt{k}}$ and to $\frac{1}{k}$.
In order to get them, one has to take into account corrections to the
saddle point method, which means that one uses Eq.\ (\ref{e2.3}) and
retains terms like those proportional to $m^2$ which give a contribution
when averaging them with the Gaussian.
It turns out, however, that these corrections are proportional
to $\frac{1}{C}$ and $\frac{1}{C^2}$, respectively, for $C<<1$, i.e.,
the correction terms diverge as the eigenvalue approaches $1$
(which is the most interesting situation for us). On the other
hand, the 'improved' solution shows a remarkable feature,
namely, that for small $C$ it depends on $C$ and on $k$ only
through the combination $C^2 k$:
\begin{eqnarray}
a^{(i)}_k=h(C^2 k)\quad.\label{e2.7}
\end{eqnarray}
If one demands this dependence
from the beginning and assumes that $C<<1,\quad k>>1$ while no assumption
is made about $C^2k$, then one gets up to order $\frac{1}{j}$
in $j$ and to order $C^2$ in $C$ (defined now through
Eq.\ (\ref{e2.6})) the equation
\begin{eqnarray}
h^{''}(z)+\frac{1}{4z}h(z)=0\quad,\label{e2.8}
\end{eqnarray}
($z$ standing for $C^2 j$)
whose two linearly independent solutions are (cf.
\cite{Abram1})
\begin{eqnarray}
h^{(1)}(z)=\sqrt{z}J_1\left(\sqrt{z}\right)\label{e2.9}
\end{eqnarray}
and
\begin{eqnarray}
h^{(2)}(z)=\sqrt{z} Y_1\left(\sqrt{z}\right)\quad,\label{e2.10}
\end{eqnarray}
$J_1$ and $Y_1$ standing for the first order Bessel and Neumann functions,
respectively. Eq.\ (\ref{e2.8}) is identical with the radial Schr\"odinger 
equation for $s$-wave scattering in a Coulomb potential at zero energy.
As discussed above, it is valid far from the 'chaotic core' of the 
effective
scattering potential which for small values of $k$ gives rise to corrections. 
We shall see, however, that the range of validity of Eq.\ (\ref{e2.8}) extends down 
to $k=1$ when $C \rightarrow 0$.
The boundary conditions follow from the restrictions that the 
eigenfunctions in the original '$x$-space' should have an integrable 
singularity at $x=0$ and their integral over the whole $[0,1]$ interval 
should vanish. This latter follows from the orthogonality of the left 
and right eigenfunctions (belonging to different eigenvalues) and from 
the fact that the identically $1$ function is the left eigenfunction
of the Frobenius-Perron operator for $C=0$ (i.e., for $\lambda=1$). Note 
that it is valid only in the case of permanent chaos and does not apply 
for transient chaos.
The eigenfunction of the Frobenius-Perron operator
is given by
\begin{eqnarray}
\phi_C(x)=\sum_{j=0}^{\infty}g_j (1-x)^{4j+1}\quad,\label{e2.11}
\end{eqnarray}
where the asymptotical form of $g_j$ is given by some linear combination
of $h^{(1)}$ and $h^{(2)}$ (see Eqs.(\ref{e2.9}),(\ref{e2.10})). As the
asymptotical form of both the Bessel and the Neumann function is
a phase-shifted sinus with a one-over-square-root type amplitude, for fixed
$C$ and increasing $j$ the moduli of the coefficients $g_j$ grow
like $j^{\frac{1}{4}}$. This growing is, however, superimposed by
the exponential decay of the $(1-x)^{4j+1}$ factor, making the
infinite sum in Eq.\ (\ref{e2.11}) absolutely convergent for $0<x\le 1$. One can 
even show that the eigenfunction $\phi_C(x)$ is integrable near zero
(one estimates the summation with an integral,
after a term-by-term integration over $x$). This is true essentially because 
$\sum_{j=0}^{\infty} g_j/(4j+2)$ is finite. One can express the 
requirement of the integrability of the eigenfunction near zero as a 
boundary condition to Eq.\ (\ref{e2.8}) stating that $\int_{1}^{\infty}h(z)/z$ is 
finite (the lower limit of the integration range being an arbitrary positive 
value).  Thus we can say that both propagating solutions (\ref{e2.9}) 
and (\ref{e2.10}) of 
Eq.\ (\ref{e2.8}) are allowed by the boundary condition at the infinity, as 
 is usually the case in the customary scattering problems. 
 
The next issue is how one can determine the actual eigenvector
$g_j$ for a given (small) $C$ knowing the asymptotical solutions
$h^{(i)}(C^2 j)$. The asymptotical solutions are actually rather
accurate approximations even for small $j$-s when $C$ is small. Nevertheless,
at or near $j=0$ they do not satisfy the eigenvalue equation,
thus giving rise to a correction term. If one starts with the
proper linear combination of the two asymptotic solutions, then
the correction term rapidly decays. This is the condition which
selects the proper asymptotics. (If, however, not the proper
linear combination has been chosen, then the correction term
itself contains an asymptotic part, describing a 'reflection' at
the origin $j=0$.) The numerical procedure
is outlined and discussed in Appendix A.
According to those considerations, for small $C$, i.e., for an eigenvalue 
near unity the eigenfunction has
the form
\begin{eqnarray}
\phi_C(x)=2(1-x)-2\sum_{j=0}^{\infty} C \sqrt{j}J_1\left(C \sqrt{j}\right)
(1-x)^{4j+1}
\quad.\label{e2.19}
\end{eqnarray}
(Here we have multiplied by $-2$ in order to exhibit the similarity to 
the $C=0$ case when just the first term of the r.h.s. of Eq.\ (\ref{e2.19}) 
remains.)
The sum may be evaluated by replacing it with
an integral. Neglecting terms proportional to $C^2$ times a
nonsingular function (which come from the derivatives of the
summand at $j=0$ when the Euler-Maclaurin Summation
Formula is applied) one obtains (cf. \cite{Abram2})
\begin{eqnarray}
\phi_C(x)=2(1-x)-\frac{C^2(1-x)}{16 \ln^2(\frac{1}{1-x})}
\exp\left(-\frac{C^2}{16 \ln(\frac{1}{1-x})}\right)
\quad.\label{e2.20}
\end{eqnarray}
For $C\rightarrow 0$ the second term vanishes everywhere except
near $x=0$, where it has a sharp peak with unit area. As the
first term is nothing but the normalized stationary probability density
of the map (which is at the same time the eigenfunction of the
Frobenius-Perron operator with unit eigenvalue (i.e., $C=0$),
one can see in which sense the limiting eigenfunction is
approached and also, that the integral of the eigenfunction over
$x$ vanishes.

One can see now that the spectral properties around the upper edge of 
the spectrum depend predominantly on the laminar motion of the map near 
its marginally unstable fixed point, as the influence of the other parts 
of the map 
can show up itself through correction terms, which, however, become 
negligible when the eigenvalue approaches unity. On the other hand, 
corrections for larger values of $C$ can be calculated with relatively 
little numerical efforts as the nondecaying asymptotical part of the 
eigenvectors (which has been previously the root of the numerical 
difficulties) is already taken into account analytically.

\section {The asymptotical solution of the Frobenius-Perron eigenvalue equation near intermittency}

We seek for the asymptotical form of the matrix elements 
$H_{j,k}$ of the 
Frobenius-Perron operator $\hat H$ corresponding to the family of maps (\ref{e1.1}) on the basis 
\begin{eqnarray}
\zeta_n\left(x\right)=\left(f_l^{-1}\right)'(x)\left(1-2\,f_l^{-1}(x)\right)^{2\,n}\nonumber\\
=\left(\frac{r+1}{2}-r\,x\right) \left(1-x\right)^{2\,n}\left(1-r\,x\right)^{2\,n}\quad 
(n=0,\,1,\,2,\,...)\label{e3.0}
\end{eqnarray}
This form is suggested by the symmetry of the map as well as by the
structure of the Frobenius-Perron operator (\ref{e1.0}).   
Indeed, for a symmetric fully developed chaotic map defined on the interval
$[0,1]$ the Frobenius-Perron operator can be expressed as 
\begin{eqnarray}
\left(\hat H \varphi\right)(x)=\left(f_l^{-1}\right)'(x)\left[\varphi\left(\frac{1}{2}-
\left(\frac{1}{2}-f_l^{-1}(x)\right)\right)+\varphi\left(\frac{1}{2}+
\left(\frac{1}{2}-f_l^{-1}(x)\right)\right)\right]\,,\label{e3.01}
\end{eqnarray}
which implies that any polynomial $\varphi(x)$ goes over under the application of
$\hat H$ into a linear combination of the functions $\left(f_l^{-1}\right)'(x)\left(1-2\,f_l^{-1}(x)\right)^{2\,n}$ which, if they are themselves polynomials, constitute a suitable basis. In our concrete example they coincide with those given above (Eq.\ (\ref{e3.0})). One can see that this basis goes over for ${r\rightarrow 1}$ to that
used in the intermittent situation.

Introducing the functions
\begin{eqnarray}
h_k\left(x\right)=\left(\hat H \zeta_k\right)\left(x\right)\:,\label{e3.1}
\end{eqnarray}
the matrix elements $H_{j,k}$ are defined by 
\begin{eqnarray}
h_k\left(x\right)=\sum_{j=0}^\infty H_{j,k} \zeta_j\:.\label{e3.2}
\end{eqnarray}
As one has to do here with polynomials, a numerical evaluation of the matrix 
elements for any given indices $j,k$ is not difficult, however, unlike in the intermittent
situation, a closed analytical expression (a counterpart of Eq.\ (\ref{e2.2}))
 this time does not exist. Nonetheless an asymptotical expression can be derived. Our starting point now is Eq.\ (\ref{e3.2}), where we insert the expressions of the basis
functions (\ref{e3.0}). Introducing the variable $z=1-2\,f_l^{-1}(x)=(1-x)(1-r\,x)$ we get
\begin{eqnarray}
2^{-(4k+1)}\left(1+r\,z\right)\left(1+z\right)^{2\,k}\left(2-r+r\,z\right)^{2\,k}\nonumber\\
+2^{-(4k+1)}\left(1-r\,z\right)
\left(1-z\right)^{2\,k}\left(2-r-r\,z\right)^{2\,k}\\ \label{e3.2a}
=
\sum_{j=0}^{\infty} H_{j,k}\,z^{2\,j}\,.\nonumber 
\end{eqnarray}
The exact matrix elements may be calculated by comparing the coefficients
of the polynomials in both sides. The result is shown in Fig. 3.

In order to derive an asymptotical analytical approximation, 
we extend the variable $z$ to the whole complex plane
and determine the matrix elements by Cauchy's formula as
\begin{eqnarray}
 H_{j,k}=\frac{1}{2\pi i}\oint \tilde h_k(z)\,z^{-(2\,j+1)}\,,\label{e3.2b}
\end{eqnarray}
where the integration contour encircles the origin and $\tilde h_k(z)$ stands for the left hand side of Eq.\ (\ref{e3.2a}). For large
$j$ and $k$ values the integrals in Eq.\ (\ref{e3.2b}) (each involving one of the terms of $\tilde h_k(z)$) may be evaluated by the saddle point method. It turns out that for a given $j$ the matrix elements have a maximum versus $k$, namely
\begin{eqnarray}
H_{j,k}\approx \frac {x}{\sqrt {\pi }}\sqrt{\frac {\left (1+r\right )^{3}}{2(1+2\,r-\,{r}^{2})}}\exp\left(-\frac {\left (1+r\right )^{3}}{2(1+2\,r-\,{r}^{2})}\left(\frac{k-\frac {2}{1+r}j}{\sqrt{j}}\right)^{2}\right)\,.
\label{e3.2c}
\end{eqnarray}
It has been assumed that $\frac{k-{\frac {2}{1+r}}
j}{\sqrt{j}}$ is of order unity (i.e., the expression (\ref{e3.2c}) is
valid near the maximum of the matrix elements in a strip of width $\approx \sqrt{j}$). 
 The most important difference between the intermittent and nonintermittent situations is that the maximal matrix
elements lie now above the diagonal, and, when increasing the index $j$, their distance from the diagonal increases
faster (linearly), than the spread of the significant matrix elements around the maximum (square root type increase). As a consequence, as we shall see, the eigenvectors
$a_k$ decay for large $k$ exponentially. Therefore, we have to evaluate Eq.\ (\ref{e3.2b}) not along $k\approx \frac {2}{1+r} j$, as we would expect from Eq.\ (\ref{e3.2c}),
but along another line $k\approx b\,j$. The quantity $b$ as well as the asymptotics $a_k\propto \exp\left({-\alpha\,k}\right)$ of the eigenvector is to be determined 
from the condition  
\begin{eqnarray}
\sum_k H_{j,k}\exp\left({-\alpha\,k}\right)\propto \exp\left({-\alpha\,j}\right)\,,
\end{eqnarray} 
where the dominant terms of the sum come from the $k$ values near $b\,j$.
Explicitly, we get 
\begin{eqnarray}
b=\frac{2}{3-r}\label{s3.1}
\end{eqnarray} 
and
\begin{eqnarray}
\alpha=2\ln\left({\frac{2-r}{r}}\right)\,.\label{s3.2}
\end{eqnarray} 
Evaluating the expression of the matrix element $H_{j,k}$ by the saddle point method
around $k= b\,j$ in a strip of width $\sqrt{j}$, we get

\begin{eqnarray}
H_{j,k}\nonumber\\
={\frac {x}{\sqrt {\pi }}}\sqrt{\frac {\left (3-r\right )^{3}}{2(1+2\,r-{r}^{2})}}\exp\left(-{2\,\frac{1-r}{3-r}\ln\left({\frac {2}{r}-1}\right)\frac{1}{x^2}
+2\,p\ln\left({\frac {2}{r}-1}\right) 
\frac{1}{x}-{\frac {\left (3-r\right )^{3}{p}^{2}}{2\,(1+2\,r-{r}^{2})}}}\right)
\label{s3.3}\\
\times \left(1+\left (b_{1,1}\,p+b_{1,3}\,p^{3}\right )x
+\left (b_{2,0}+b_{2,2}\,p^{2}+b_{2,4}\,p^{4}+b_{2,6}\,p^{6}
\right )x^{2}\right )\nonumber
\end{eqnarray}
where
\begin{eqnarray}
b_{1,1}={\frac {\left ({r}^{4}-12\,{r}^{3}+36\,{r}^{2}-28\,r-9\right )\left (
3-r\right )}{4\,\left (1+2\,r-{r}^{2}\right )^{2}}}\label{s3.4}
\end{eqnarray}
\begin{eqnarray}
b_{1,3}={\frac {\left ({r}^{4}-16\,{r}^{2}+24\,r+3\right )\left (3-r\right )
^{4}}{12\,\left (1+2\,r-{r}^{2}\right )^{3}}}\label{s3.5}
\end{eqnarray}
\begin{eqnarray}
b_{2,0}={\frac {{r}^{7}-33\,{r}^{6}+235\,{r}^{5}-687\,{r}^{4}+835\,{r}^{3}-
171\,{r}^{2}-243\,r-57}{48\,\left (1+2\,r-{r}^{2}\right )^{3}}}\label{s3.6}
\end{eqnarray}
\begin{eqnarray}
b_{2,2}=-{\frac {\left ({r}^{8}+8\,{r}^{7}-240\,{r}^{6}+1496\,{r}^{5}-4010\,{
r}^{4}+4664\,{r}^{3}-1216\,{r}^{2}-952\,r-183\right )\left (3-r
\right )^{2}}{32\,\left (1+2\,r-{r}^{2}\right )^{4}}}\label{s3.7}
\end{eqnarray}
\begin{eqnarray}
b_{2,4}=-{\frac {\left (4\,{r}^{7}-9\,{r}^{6}-122
\,{r}^{5}+629\,{r}^{4}-1040\,{r}^{3}+459\,{r}^{2}+186\,r+21\right )\left (3-r\right )^{5}}{24\,
\left (1+2\,r-{r}^{2}\right )^{5}}}\label{s3.8}
\end{eqnarray}
\begin{eqnarray}
b_{2,6}={\frac {\left (3-r\right )^{8}\left ({r}^{4}-16\,{r}^{2}+24\,r+3
\right )^{2}}{288\,\left (1+2\,r-{r}^{2}\right )^{6}}}\label{s3.9}
\end{eqnarray}
and
\begin{eqnarray}
x=\frac{1}{\sqrt{j}}\label{s3.10}
\end{eqnarray}
\begin{eqnarray}
p=\frac{k-\frac{2}{3-r}j}{\sqrt{j}}\,.\label{s3.11}
\end{eqnarray}

Figs. 4, 5 show the differences between the 
approximate expressions (\ref{e3.2c}), (\ref{s3.3})
and the exact matrix elements.

Our next task is to solve the eigenvalue equation
\begin{eqnarray}
\sum_{k=0}^{\infty} H_{j,k} s_k=\lambda s_j\,.\label{e3.2e}
\end{eqnarray}
We approximate the summation by an integral (considering $k$ to be a continuous variable), extend its lower limit to $-\infty$, insert the asymptotical expression (\ref{e3.2c}) for the matrix elements and evaluate the left hand side by the saddle point method. 
In order to make it systematically, we write 
\begin{eqnarray}
s_k=\varphi(k)\exp\left({-\alpha\,k}\right)\,,\label{e3.2f}
\end{eqnarray}
and assume that $\varphi(k)$ grows (or decays)  asymptotically at most like a power of $k$. Then we insert the expression (\ref{s3.3}) of $H_{j,k}$, expand $\varphi(k)$ around
$b\,j$ on the l.h.s. up to fourth order, and perform the integration. The result 
can be written down for arbitrary $r$, but for simplicity it will be presented 
for $r=1-\epsilon$, where $\epsilon<<1$, i.e., near the intermittent situation.
As the l.h.s. of the equation becomes a function of $b\,j$, while at the r.h.s.  
$\varphi(j)$ stands, the latter should also be expanded around $b\,j$. Introducing
the new independent variable
\begin{eqnarray}
\tilde x=4\epsilon\, b\,j\,\label{s3.12} 
\end{eqnarray}
and writing
\begin{eqnarray}
\varphi(b\,j)=\xi(\tilde x)\,,\label{s3.13}
\end{eqnarray} 
we arrive at the equation
\begin{eqnarray}
\tilde x\xi''-\tilde x\xi'+\frac{2(1-\lambda)}{\epsilon\lambda}\xi\nonumber\\
+\epsilon\left[\frac{1}{4}\tilde x^2\xi''''
+\tilde x\xi'''+\left(-\frac{\tilde x^2}{4}
-\tilde x+2\right)\xi''-2\xi'\right]=0\,.\label{s3.14}
\end{eqnarray}
For small $\epsilon$ the term proportional to $\epsilon$ may be neglected in the
first approximation. The resulting equation,
\begin{eqnarray}
\tilde x\xi''-\tilde x\xi'+\frac{2(1-\lambda)}{\epsilon\lambda}\xi=0\label{s3.14a}
\end{eqnarray}
 has a solution which diverges
for $\tilde x\rightarrow \infty$ slower than exponentially (actually, as a power) only
if  $\frac{2(1-\lambda)}{\epsilon\lambda}$ is a positive integer $n$. In this case
\begin{eqnarray}
\xi=\xi_n=\tilde x\,L^{1}_{n-1}(\tilde x)\,,\label{s3.15}
\end{eqnarray}
and the eigenvector $s_j^{(n)}$ (cf. Eqs.(\ref{e3.2f})-(\ref{s3.13}))
is expressed as
\begin{eqnarray}
s_j^{(n)}\,=\,4\epsilon\, b\,j\,L^{1}_{n-1}(4\epsilon\, b\,j)\,
\exp\left(-\alpha\,j\right)\,,\label{s3.15e}
\end{eqnarray}
where $L^{1}_{n-1}(\tilde x)$ stands for the generalized Laguerre polynomial\cite{Abram3a}.
The eigenvalue $\lambda$ is given (up to first order in $\epsilon$) by
\begin{eqnarray}
\lambda_n=1-\frac{\epsilon}{2}n\,.\label{s3.16}
\end{eqnarray}
Let us consider the transition to the intermittent case.
Eq.\ (\ref{s3.16}) implies that in that case (i.e., when $\epsilon
\rightarrow 0$) the spacing between neighbouring eigenvalues
vanishes, thus we get a continous spectrum. In order to get the eigenvectors,
we fix the value of $\lambda$, thus also of $\epsilon n$ and then
take the limit $\epsilon \rightarrow 0$, or, equivalently,
$n \rightarrow \infty$. We obtain \cite{abramx}
\begin{eqnarray}
s^{\lambda}_j=\lim_{n\rightarrow \infty}
\frac{8(1-\lambda)bj}{n}L^{(1)}_{n-1}\left(\frac{8(1-\lambda)bj}{n}\right)
e^{-\alpha j}=\sqrt{8(1-\lambda)j}\,J_1\left(2\sqrt{8(1-\lambda)j}\right)
\end{eqnarray}
Provided that $1-\lambda << 1$, we may write according to Eq.\ (\ref{e2.6})
$1-\lambda\approx \frac{C^2}{32}$, thus we get
\begin{eqnarray}
s^{\lambda}_j\approx \frac{C}{2}\sqrt{j}\,J_1\left(C\sqrt{j}\right)\,,
\end{eqnarray}
in accordance with Eq.\ (\ref{e2.9}), which gives the dominant
contribution near the upper edge of the spectrum (cf. the discussion
after Eq.\ (\ref{e2.10})).

When deriving Eqs.\ (\ref{s3.15})-(\ref{s3.16}) we have made 
use of the same argument as that applied at the solution
of the radial Schr\"odinger equation with a Coulomb potential. It is indeed possible
to cast Eq.\ (\ref{s3.14a}) to the same form. To do this, let us introduce
the new independent variable 
\begin{eqnarray}
\rho=\kappa \tilde x\label{s3.17}
\end{eqnarray}
and the new function
\begin{eqnarray}
\chi(\rho)=\xi \exp\left(-\frac{\rho}{2\,\kappa}\right)\,,\label{s3.18}
\end{eqnarray}
where
\begin{eqnarray}
\kappa=\frac{2(1-\lambda)}{\epsilon\lambda}\,.\label{s3.19}
\end{eqnarray}
Then we get from Eq.\ (\ref{s3.14a})
\begin{eqnarray}
-\chi''-\frac{1}{\rho}\chi=-\frac{1}{4\,\kappa^2}\chi\,.\label{s3.20}
\end{eqnarray}
This is the well-known form of the radial Schr\"odinger equation. Note that the
actual radial wave function $R(\rho)$ corresponds to $\chi /\rho$. The term 
 $-\frac{1}{\rho}$ represents an attracting Coulomb potential and 
$E=-\frac{1}{4\,\kappa^2}$ is the Rydberg-formula. As usually, bound states 
are associated with positive integer values for $\kappa$. 
We may apply the transformation (\ref{s3.17})-(\ref{s3.18}) also to Eq.\ (\ref{s3.14}). 
Before doing that we simplify somewhat the correction term. We shall be interested 
only in the next order correction to the eigenvalue (\ref{s3.16}), thus
in the order $\epsilon$ term of Eq.\ (\ref{s3.14}) we may express all the higher 
derivatives in terms of $\xi'$ and of $\xi$, as we may assume at the given 
accuracy that $\xi$ satisfies Eq.\ (\ref{s3.14a}). Thus we get for the $n$-th
eigenvalue and eigenfunction
\begin{eqnarray}
\tilde x\xi''-\tilde x\xi'+\kappa \xi+\epsilon\left[\left(\frac{n(n+1)}{4}
-\frac{3n}{2\tilde x}\right)\xi -\frac{n}{2}(1+\tilde x)\xi'\right]=0\,.\label{s3.21}
\end{eqnarray}
Performing now the transformation (\ref{s3.17})-(\ref{s3.18}), we get
\begin{eqnarray}
-\chi''-\frac{1}{\rho}\chi
-\epsilon\left[\left(-\frac{3n}{2}\frac{1}{\rho^2}+\frac{n}{4}\frac{1}{\rho}
-\frac{1}{4n}\right)\chi-\left(\frac{n}{2}\frac{1}{\rho}+\frac{1}{2}\right)\chi'\right]
=-\frac{1}{4\,\kappa^2}\chi\,.\label{s3.21a}
\end{eqnarray}
Applying first order perturbation theory, we may express the first correction
of $-\frac{1}{4\,\kappa^2}$ as the diagonal matrix element of the perturbing
operator, i.e., 
\begin{eqnarray}
\delta\left(-\frac{1}{4\,\kappa^2}\right)
=-\epsilon\int_0^\infty d\rho\, \chi_n(\rho)\left[\left(-\frac{3n}{2}\frac{1}{\rho^2}
+\frac{n}{4}\frac{1}{\rho}-\frac{1}{4n}\right)\chi_n(\rho)-\left(\frac{n}{2}\frac{1}{\rho}
+\frac{1}{2}\right)\chi_n'(\rho)\right]\,.\label{s3.22}
\end{eqnarray}
Using Eq.\ (\ref{s3.19}) this leads to the expression of the eigenvalue $\lambda$
\begin{eqnarray}
\lambda=1-\frac{n}{2}\epsilon-\left(\frac{7}{8}n-\frac{1}{8}n^2\right)\epsilon^2+
O(\epsilon^3)=\exp\left[-n\left(\frac{\epsilon}{2}+\frac{7}{8}\epsilon^2\right)\right]+
O(\epsilon^3)\,,\label{s3.23}
\end{eqnarray}
which is valid in $\epsilon$ up to second order, provided that not only
$\epsilon$, but also $\epsilon n$ is much smaller than $1$.
A comparison of these results with those of a direct numerical
measurement \cite{KLB} is displayed in Fig. 6. 

As for a comparison in case of
the eigenfunctions, first we have to determine them in coordinate	 
space. Eqs.\ (\ref{s3.15e}) and (\ref{e3.0}) imply
\begin{eqnarray}
s^{(n)}(x)=\sum_{j=0}^\infty 4\epsilon b j L_{n-1}^{(1)}(4\epsilon b j)
\exp(-\alpha j)\left(\frac{r+1}{2}-rx\right)
(1-x)^{2j}(1-rx)^{2j}\nonumber\\
\approx \frac{1}{4\epsilon b}\left(\frac{r+1}{2}-rx\right)\int_0^\infty
\,dy\,yL_{n-1}^{(1)}(y)e^{-\kappa (x)y}\,,\label{sfv}
\end{eqnarray}
where
\begin{eqnarray}
\kappa (x)=\frac{1}{4\epsilon b}\left[\alpha 
- 2\ln\left((1-x)(1-rx)\right)\right]\,.\label{kpx}
\end{eqnarray}
Using Rodrigues' formula \cite{abramy}
\begin{eqnarray}
L_{n}^{(1)}(x)=\frac{1}{n!}\frac{e^x}{x}\frac{d^n}{dx^n}
\left(x^{n+1}e^{-x}\right)
\end{eqnarray}
we may evaluate the integral in (\ref{sfv}) to get
\begin{eqnarray}
s^{(n)}(x)=\frac{n}{4\epsilon b}\left(\frac{r+1}{2}-r x\right)
\frac{(\kappa(x) -1)^{n-1} }{\kappa(x)^{n+1} }\,.\label{sfv1}
\end{eqnarray}
As our asymptotical method neglects any effects which appear
at $j=0$, the integral of these functions does not vanish. Therefore,
a suitable multiple of $P(x)=2\zeta_0(x)$ must be still subtracted.
The resulting expression, however, gives a reasonable fit
only quite near to the intermittent case $r=1$, so that for $r<0.99$ 
one should calculate corrections as well. The essence of the
 problem is not to 
push perturbation theory to higher order, but
to take into account the second independent solution of the
equation (\ref{s3.14}) and to add corrections near the origin,
where the asymptotic expansion does not hold any longer. The
procedure is in complete analogy with the intermittent situation (cf. Appendix A).
It may be surprising that the second independent solution also
plays a role. Indeed, at this point the analogy with the quantum mechanical
case breaks down. The reason is that the physical meaning of the 
eigenfunctions is different, especially, they are differently
related to probability distributions.
Therefore in the quantum mechanical case a singularity in the
origin is not allowed, while it is allowed in our case.
Note that both solutions correspond to the same eigenvalue,
i.e., the boundary condition at infinity completely determines
the spectrum.
The calculation of the second type of eigenfunctions is simpler
after a Laplace transform, which brings us back to the original
coordinate $x$. Indeed,
\begin{eqnarray}
s^{(n)}(x)=\sum_{j=0}^\infty \xi_n(4\epsilon b j){\rm e}^{-\alpha j}
\zeta_j(x)\nonumber\\
=\sum_{j=0}^\infty \xi_n(4\epsilon b j){\rm e}^{-\alpha j}\left(\frac{r+1}{2}
-rx\right)(1-x)^{2j}(1-rx)^{2j}\\
\approx\left(\frac{r+1}{2} -rx\right)\frac{1}{4\epsilon b}
\int_0^\infty dz \xi_n(z) {\rm e}^{-\kappa(x) z}\nonumber
\end{eqnarray}
where $\kappa(x)$ is given by Eq.\ (\ref{kpx}).
Taking the Laplace transform of Eq.\ (\ref{s3.14}) we get 
\begin{eqnarray}
\xi_n(0)+(n+1-2 \kappa)\tilde \xi(\kappa)-\kappa(\kappa-1)\frac{d\tilde \xi(\kappa)}{d\kappa}=0\,,\label{ltq}
\end{eqnarray}
where
\begin{eqnarray}
\tilde \xi(\kappa)=\int_0^\infty dz \xi_n(z) {\rm e}^{-\kappa z}
\end{eqnarray}
stands for the Laplace transform of $\xi_n(z)$. 
Choosing $\xi_n(0)=0$ in Eq.\ (\ref{ltq}) and substituting (\ref{kpx}) we
arrive at the 'regular' eigenfunctions (\ref{sfv1}) again. The 'irregular'
(logarithmic) eigenfunctions correspond to the choice $\xi_n(0)\ne 0$.
Choosing $\xi_n(0)=-1$ we get
\begin{eqnarray}
\tilde \xi(\kappa)=-\frac{(\kappa -1)^n}{\kappa^{n+1}}-n\ln(\kappa -1)
\frac{(\kappa -1)^{n-1}}{\kappa^{n+1}}
+\sum_{j=2}^n \frac{1}{j-1}{n \choose j} 
\frac{(\kappa -1)^{n-j}}{\kappa^{n+1}}\,.
\end{eqnarray}
Taking the inverse Laplace transform we get the second independent
solution for $\xi_n(z)$. It 
can be given as an infinite sum, but we do not
reproduce here this cumbersome expression. 

In order to get a reasonable approximation for the eigenfunctions
even for relatively small $r$ values, we apply the scheme
decribed in Appendix A (cf. Eqs.\ (\ref{e2.12})-\ (\ref{e2.16})). 
Now the two independent 
asymptotical solutions $h^{(1)}_j$, $h^{(2)}_j$ 
correspond to the regular and irregular solutions for $\xi_n(z)$,
and the exact matrix elements $H_{j,k}$ 
valid in the nonintermittent case (cf. Eq.\ (\ref{e3.2a})) should
be used. As a demonstration, 
in Figs. 7-9
the results 
for the first three eigenfunctions 
at $r=0.95$ 
are compared with those 
of the numerical measurements. In this case the corrections played 
already an important role, $37$ terms had to be included. 
We also found that it was essential to use
good approximations for the eigenvalues (cf. Eq.\ (\ref{s3.23})).    

\section{Correlation functions}

Recalling the definition (\ref{e1.0}) 
of the Frobenius-Perron operator $\hat H$,
one may cast a correlation function
\begin{eqnarray}
C_t^{A,B}=\int_0^1 \;dx\;B\left(f^{[t]}(x)\right)\;A(x)P(x)\label{c1}
\end{eqnarray}
($f^{[t]}(x)$ standing for the $t$-th iterate of the mapping $f(x)$)
to the form
\begin{eqnarray}
C_t^{A,B}=\int_0^1 \;dx\;B(x)\;\hat H^t\left(A(x)P(x)\right)i\,.\label{c2}
\end{eqnarray}
Expanding now $A(x)P(x)$ in terms of the eigenfunctions of the Frobenius-Perron
operator the action of $\hat H^t$ reduces to a multiplication of each term
by the $t$-th power of the corresponding eigenvalue. The details of the 
calculation outlined here are presented in Appendix B. There we make
use of the previously introduced infinite matrix representation (\ref{e3.2})
and apply Eq.\ (\ref{s3.15e}) for the eigenvectors. This is allowed
under the assumption that we are close to an intermittent situation
(i.e., corrections and the irregular Coulomb functions may be 
neglected) and also that the correlators
$A(x)$ and $B(x)$ have already zero mean with respect to the natural
measure, therefore, a substraction of a multiple of $P(x)$ from the
eigenfunctions is not necessary. Then we apply identities for the
generalized Laguerre polynomials, substitute sums with integrals (which
is made possible again by the closeness of the intermittent situation),
and finally end up with the expression  
\begin{eqnarray}
C_t^{A,B}\approx \int_0^1 dx\,P(x)\,B(x)\,
\frac{\tau \left(4\epsilon b\right)^2}{\left[4\epsilon b\tau 
+(1-\tau)\left(\alpha-2\ln\left((1-x)(1-rx)\right)\right)\right]^2}
\label{c3}\\
\times A\left(f\left(\frac{1}{2}-\frac{1}{2}\exp\left(-\frac{1}{2}
\frac{\alpha(1-\tau)\left(4\epsilon b -\alpha\right)
-2\ln\left((1-x)(1-r x)\right)\left(4\epsilon b -\alpha(1-\tau)\right)}
{4\epsilon b\tau +(1-\tau)\left(\alpha-2\ln\left((1-x)(1-r x)\right)\right)}
\right)\right)\right)\,.\nonumber
\end{eqnarray}
Here 
\begin{eqnarray}
\tau=\exp\left(-\left(\frac{\epsilon}{2}+\frac{7}{8}\epsilon^2\right)t\right)\\
\epsilon=1-r\\
\alpha=2\ln\left(\frac{2-r}{r}\right)\\
b=\frac{2}{3-r}
\end{eqnarray}
A comparison with the results of a direct numerical measurement of the
correlation function is displayed in Fig. 10.
 (cf. Fig.1. in Ref.\cite{LBK},
where the same is displayed with a rougher analytical estimate).
The functions $A(x)$ and $B(x)$ are given by
\begin{eqnarray}
A(x)=B(x)=\left\{\begin{array}{cr}
1&\mbox{\rm if }\,x< 5\times 10^{-2}\\
0&\mbox{\rm otherwise}
\end{array}\right.\label{c4}
\end{eqnarray} 

\section{Conclusion}

A new analytical method has been developed for the determination
of spectral properties near and at the intermittent situation
of fully developed chaotic one-dimensional maps. We made use of an infinite
dimensional matrix representation, determined the asymptotical
expression of the matrix elements and reduced the eigenvalue equation
to a differential equation, which in the lowest order coincides
with the radial Schr\"odinger equation of the hydrogen atom for
$s$-states. This allows for the calculation of corrections
by using the standard quantum mechanical perturbation theory.
We have demonstrated this by calculating the correction
to the eigenvalue. The eigenfunctions were also analytically
determined and compared with the results of numerical measurements.
Finally, we derived an analytical expression for the
correlation functions. It has been shown to fit well the numerical data.
 
One may think that our study applies only to a very specific
example. Actually, the method may be applied to any fully developed
chaotic one-dimensional map which is near to an intermittent situation
and which has analytic inverse branches. 
The universal character of our results
(within the class of mappings mentioned above) has been discussed in
 Ref.\ \cite{LBK}.
We expect that the method
can be extended to the case of repellers, where one also has 
a complete symbolic dynamics. As a motivation we mention that
in case of repellers a special
interplay of transient chaos and intermittency appears,
as discussed in the recent papers \cite{LNSz},
\cite{KLNSz} which the interested reader may consult.

\begin{appendix}
\section{Calculating corrections in the intermittent situation}

The numerical procedure is the following:
one assumes that for some $n$ the correction term $\delta g_j$
is negligible when $j>n$, and then determines both the proper
asymptotics and $\delta g_j$ from the requirement that the sum
\begin{eqnarray}
\sum_{j=n/2+1}^n (\delta g_j)^2\label{e2.12}
\end{eqnarray}
be minimal (the elements in the second half of the correction
vector are involved). Explicitely, it gives
\begin{eqnarray}
\delta g_j=\cos\left(\alpha\right)\delta^{(1)}_j+\sin\left(\alpha\right)\delta^{(2)}_j\label{e2.13}
\end{eqnarray}
and the proper asymptotics is given by
\begin{eqnarray}
\cos\left(\alpha\right)h^{(1)}_j+\sin \left(\alpha \right)h^{(2)}_j\quad,\label{e2.14}
\end{eqnarray}
the accurate eigenvector $g_j$ being the difference of the above two
quantities. Here $\delta^{(i)}_j$ is the solution of the equation
\begin{eqnarray}
\sum_{k=0}^n H_{j,k}\delta^{(i)}_k - \lambda \delta^{(i)}_j=
\sum_{k=0}^{\infty} H_{j,k}h^{(i)}(C^2 k) - \lambda h^{(i)}(C^2 j)\quad.\label{e2.15}
\end{eqnarray}
These are the first $n$ of the set of the eigenvalue equations.
Note that if the correction vector $\delta g_j$ is of length $n$
(as is assumed here), its last $n/2$ components still enter the next $n$
equations as well. Thus, when these components are small (as
required), the error caused by them is small and the
procedure is consistent. Indeed, for $C=0.1$ and $n=10$ these
components are less than $10^{-5}$ (cf. Table 1).
The 'phase shift' $\alpha$ is given by
\begin{eqnarray}
tan(2\alpha)=\frac{2 \sum_{j=n/2+1}^n \delta^{(1)}_j \delta^{(2)}_j}{
\sum_{j=n/2+1}^n (\delta^{(1)}_j)^2-(\delta^{(2)}_j)^2}
\quad,\label{e2.16}
\end{eqnarray}
as implied by Eqs.(\ref{e2.12}), (\ref{e2.13}). 
Numerically we get, as Table 1 demonstrates, that in case of
$C\rightarrow 0$ the 'phase shift' vanishes and $\delta
g_j\rightarrow -\delta_{0,j}$. This result can be understood as follows.
For $C<<1$ and for $0<j<<\frac{1}{C^2}$ the asymptotical solutions 
$h^{1,2}(z)$ can be written approximately as
\begin{eqnarray}
h^{(1)}(z)\approx \frac{1}{2}C^2 j \nonumber\\
h^{(2)}(z)\approx -\frac{2}{\pi}\quad,\label{e2.17}
\end{eqnarray}
hence the eigenvector $g_j$ is of the form
\begin{eqnarray}
g_j=\frac{1}{2}C^2 j \cos{\alpha}-\frac{2}{\pi}\sin{\alpha}+\delta 
g_j\quad.\label{e2.18}
\end{eqnarray} 
As the matrix elements $H_{j,k}$ are of order unity (compared to $C$), 
if $\alpha=O(C^2)$, then $\delta g_j$ will be also of order $C^2$. For 
$j>\frac{1}{C^2}$ the asymptotical solutions are already accurate to 
order $C^2$ (provided that $C$ is small enough), thus for those values 
of $j$ $\delta g_j$ practically vanishes. In view of Eq.\ (\ref{e2.15}) this means 
that the above estimates for $\alpha$ and $\delta g_j$ indeed hold. Note 
that for $j=0$ the combination $(1-\lambda)\delta g_0\approx \frac{C^2}{32} 
\delta g_0$ enters the eigenvalue equation, hence (unlike when $j>0$) 
$\delta g_0=O(1)$. Another important observation refers to the boundary 
conditions to Eq.\ (\ref{e2.8}): as we have seen, for $C<<1$ 
we get $h(z)\propto z$ (cf. 
Eqs.(\ref{e2.9}) and (\ref{e2.14})), just like in the case of the quantum mechanical 
scattering on a pure Coulomb potential.
\section{Derivation of an analytic expression for the correlation
functions}

\begin{eqnarray}
C_t^{A,B}=\int_0^1 \;dx\;B(x)\;\hat H^t\left(A(x)P(x)\right)\nonumber\\
=\int_0^1 \;dx\;B(x)\;\hat H^t\left(\sum_{j=0}^\infty a(j) \zeta_j(x)\right)
\label{s4.24}\\
=\int_0^1 \;dx\;B(x)\;\sum_{j=0}^\infty a(j) \sum_{k=0}^\infty \left(H^t\right)_{k,j}\zeta_k(x)\nonumber\\
=\sum_{j,k} \left[\int_0^1 \;dx\;B(x)\;\zeta_k(x)\right] \left(H^t\right)_{k,j}a(j) \nonumber\\
\end{eqnarray}
introducing 
\begin{eqnarray}
\beta(\tilde x)=\int_0^1 \;dx\;B(x)\;\zeta_{\frac{\tilde x}{4\epsilon b}}(x)\label{s4.26}
\end{eqnarray}
and
\begin{eqnarray}
G_t(\tilde x,\tilde y)= \frac{1}{4\epsilon b}
\left(H^t\right)_{\frac{\tilde x}{4\epsilon b},\frac{\tilde y}{4\epsilon b}}
\end{eqnarray}
we may write Eq.\ (\ref{s4.24}) as
\begin{eqnarray}
C_t^{A,B}\approx \frac{1}{4\epsilon b} \int_0^\infty d\tilde x \int_0^\infty d\tilde y \beta(\tilde x)
G_t(\tilde x,\tilde y)a\left(\frac{\tilde y}{4\epsilon b}\right)\label{s4.25}
\end{eqnarray}
Making use of the eigenfunctions Eq.\ (\ref{s3.15e}) we may write
\begin{eqnarray}
\int_0^\infty d\tilde y G_t(\tilde x,\tilde y)s^{(l)}_{\frac{\tilde y}{4\epsilon b}}
=\sum_j \left(H^t\right)_{k,j}s^{(l)}_j=\lambda_l^t s^{(l)}_{\frac{\tilde x}{4\epsilon b}}
\end{eqnarray}
Let us represent $a\left({\frac{\tilde y}{4\epsilon b}}\right)$ (cf. Eq.\ (\ref{s4.25})) as
\begin{eqnarray}
a\left({\frac{\tilde y}{4\epsilon b}}\right)=\sum_{l=1}^\infty c_l s^{(l)}_{\frac{\tilde y}{4\epsilon b}}
=\sum_{l=1}^\infty c_l\tilde y L^1_{l-1}(\tilde y)\,\exp\left(-\frac{\alpha}{4\epsilon b}\tilde y\right)
\end{eqnarray}
where
\begin{eqnarray}
c_l=\frac{1}{l}\int_0^\infty d\tilde y a\left({\frac{\tilde y}{4\epsilon b}}\right)
\exp\left(\frac{\alpha}{4\epsilon b}\tilde y\right)
\exp(-\tilde y)L^1_{l-1}(\tilde y)\;.
\end{eqnarray}
Then we may write
\begin{eqnarray}
\int_0^\infty d\tilde y G_t(\tilde x,\tilde y)a\left({\frac{\tilde y}{4\epsilon b}}\right)\nonumber\\
=\int_0^\infty d\tilde y G_t(\tilde x,\tilde y)\sum_{l=1}^\infty 
\frac{1}{l}\int_0^\infty d\tilde y' a\left({\frac{\tilde y'}{4\epsilon b}}\right)
\exp\left(-\left(1-\frac{\alpha}{4\epsilon b}\right)\tilde y'\right)L^1_{l-1}(\tilde y')\tilde y L^1_{l-1}(\tilde y)
\exp\left(-\frac{\alpha}{4\epsilon b}\tilde y\right)\nonumber\\
=\sum_{l=1}^\infty \frac{1}{l}\int_0^\infty d\tilde y' a\left({\frac{\tilde y'}{4\epsilon b}}\right)
\exp\left(-\left(1-\frac{\alpha}{4\epsilon b}\right)\tilde y'\right)L^1_{l-1}(\tilde y')\lambda_l^t 
\tilde x L^1_{l-1}(\tilde x)\exp\left(-\frac{\alpha}{4\epsilon b}\tilde x\right)\label{s4.0}\\
=\int_0^\infty d\tilde y \exp\left(-\left(1-\frac{\alpha}{4\epsilon b}\right)\tilde y\right) 
\tilde x \exp\left(-\frac{\alpha}{4\epsilon b}\tilde x\right)\sum_{l=1}^\infty L^1_{l-1}(\tilde x)
\frac{1}{l}L^1_{l-1}(\tilde y)\lambda_l^t a\left({\frac{\tilde y}{4\epsilon b}}\right)\nonumber
\end{eqnarray}
that implies
\begin{eqnarray}
G_t(\tilde x,\tilde y)=\tilde x 
\exp\left(-\frac{\alpha}{4\epsilon b}\tilde x
-\left(1-\frac{\alpha}{4\epsilon b}\right)\tilde y\right) 
\sum_{l=1}^\infty L^1_{l-1}(\tilde x)
\frac{1}{l}L^1_{l-1}(\tilde y)\lambda_l^t
\end{eqnarray}
Here $\lambda_l$ is of the form $\exp(-\gamma l)$ (cf. Eq.\ (\ref{s3.23})) where
$\gamma=\frac{\epsilon}{2}+\frac{7}{8}\epsilon^2$. Inserting this and applying the 
identity\cite{Abram3}
\begin{eqnarray}
\frac{1}{l}L^1_{l-1}(\tilde y)=\frac{1}{\tilde y}
\left(L^0_{l-1}(\tilde y)-L^0_l(\tilde y)\right)
\end{eqnarray}
we obtain
\begin{eqnarray}
G_t(\tilde x,\tilde y)=\frac{\tilde x}{\tilde y}
\exp\left(-\frac{\alpha}{4\epsilon b}\tilde x
-\left(1-\frac{\alpha}{4\epsilon b}\right)\tilde y\right) 
 \sum_{l=1}^\infty L^1_{l-1}(\tilde x)
 \left(L^0_{l-1}(\tilde y)-L^0_l(\tilde y)\right)\tau^l
\end{eqnarray}
where $\tau$ stands for $\exp(-\gamma t)$.
Let us insert the contour integral representation \cite{Abram4}
\begin{eqnarray}
L^0_l(\tilde y)=\frac{\exp(\tilde y)}{2\pi i}\oint_C dz \frac{\exp(-z)}{z-\tilde y}
\left(\frac{z}{z-\tilde y}\right)^l
\end{eqnarray}
where the contour $C$ encircles the point $z=\tilde y$. We get
\begin{eqnarray}
G_t(\tilde x,\tilde y)=-\frac{\tilde x \;\tau}{2\pi i}\exp\left(-\frac{\alpha}{4\epsilon b}
(\tilde x -\tilde y)  \right)\oint_C dz 
\frac{\exp(-z)}{(z-\tilde y)^2} \sum_{l=1}^\infty L^1_{l-1}(\tilde x)
\left(\frac{z\;\tau}{z-\tilde y}\right)^{l-1}\;.
\end{eqnarray}
Using the generating function of the Laguerre polynomials\cite{Abram5}, i.e., the identity
\begin{eqnarray}
\sum_{l=1}^\infty L^1_{l-1}(\tilde x) u^{l-1}=\frac{1}{(1-u)^2}\exp\left(\frac{\tilde x u}{u-1}\right)
\end{eqnarray}
that holds true for 
\begin{eqnarray}
|u|<1\label{cond}
\end{eqnarray}
 we get finally
\begin{eqnarray}
G_t(\tilde x,\tilde y)=-\frac{\tilde x}{\tilde y}\frac{\tau}{2\pi i}\exp\left(-\frac{\alpha}{4\epsilon b}
(\tilde x -\tilde y)  \right)\oint_{C'} dw
\frac{\exp\left(-\tilde y w -\frac{\tilde x \tau w}{w(1-\tau)-1}\right)}{(w(1-\tau)-1)^2}\;.\label{s4.27}
\end{eqnarray}
Here the new complex variable $w=z/\tilde y$ has been introduced. The contour $C'$ encircles
accordingly the points $w=1$ and $w=1/(1-\tau)$, the latter coming from the condition (\ref{cond}).
Inserting  Eqs.(\ref{s4.27}), (\ref{s4.26}), (\ref{e3.0}) into Eq.\ (\ref{s4.25})
the integration over $\tilde x$ may be performed to get
\begin{eqnarray}
C_t^{A,B}\approx 2\epsilon b\tau\int_0^1 dx\,P(x)\,B(x)\,\nonumber\\
\times\left(-\frac{1}{2 \pi i}\right)\oint_{C'} dw \frac{\tilde a\left(4\epsilon b w-\alpha\right)}
{\left[w\left[4\epsilon b\tau +(1-\tau)\left(\alpha-2\ln\left((1-x)(1-r x)\right)\right)\right]
-\left(\alpha-2\ln\left((1-x)(1-r x)\right)\right)\right]^2} \label{cr0}\\
=-8\epsilon^2 b^2\tau\int_0^1 dx\,P(x)\,B(x)\,
\tilde a'\left(
\frac{\alpha(1-\tau)\left(4\epsilon b -\alpha\right)
-2\ln\left((1-x)(1-r x)\right)\left(4\epsilon b -\alpha(1-\tau)\right)}
{4\epsilon b\tau +(1-\tau)\left(\alpha-2\ln\left((1-x)(1-r x)\right)\right)}
\right)\nonumber\\
\times\left[4\epsilon b\tau 
+(1-\tau)\left(\alpha-2\ln\left((1-x)(1-rx)\right)\right)\right]^{-2}
\,.\nonumber
\end{eqnarray}
Here we have also used the fact that in our case $P(x)=2\,\left(f_l^{-1}\right)'(x)$. Furthermore,
\begin{eqnarray}
\tilde a(p)=\int_0^\infty dy \exp\left(-y p\right) \frac{a(y)}{y}\,.
\end{eqnarray}
Consider now the expression
\begin{eqnarray}
-\tilde a'(p)=
\int_0^\infty d y \exp\left(- y p\right) a(y)
\\
\approx\sum_{j=0}^\infty a(j) 
\exp\left(-p j\right)\,.
\end{eqnarray}
Comparing this with the expansion (cf.(\ref{s4.24}))
\begin{eqnarray}
A(x)P(x)=\sum_{j=0}^\infty a(j) \zeta_j(x)
=\sum_{j=0}^\infty a(j) \frac{1}{2}P(x)\left(1-2 f_l^{-1}(x)\right)^{2j}
\end{eqnarray}
we get
\begin{eqnarray}
-\tilde a'(p)\approx
2 A\left(f\left(\frac{1}{2}\left(1-
\exp\left(-\frac{p}{2}\right)\right)\right)\right)
\end{eqnarray}
Inserting this into Eq.\ (\ref{cr0}) we get the final expression for the correlation function:
\begin{eqnarray}
C_t^{A,B}\approx \int_0^1 dx\,P(x)\,B(x)\,
\frac{\tau \left(4\epsilon b\right)^2}{\left[4\epsilon b\tau 
+(1-\tau)\left(\alpha-2\ln\left((1-x)(1-rx)\right)\right)\right]^2}
\label{kesz}\\
\times A\left(f\left(\frac{1}{2}-\frac{1}{2}\exp\left(-\frac{1}{2}
\frac{\alpha(1-\tau)\left(4\epsilon b -\alpha\right)
-2\ln\left((1-x)(1-r x)\right)\left(4\epsilon b -\alpha(1-\tau)\right)}
{4\epsilon b\tau +(1-\tau)\left(\alpha-2\ln\left((1-x)(1-r x)\right)\right)}
\right)\right)\right)\,.\nonumber
\end{eqnarray}
\end{appendix}

\newpage

\begin{table}

\begin{tabular}{|@{\extracolsep{\fill}}c|l|l|l|l|}
$\mbox{\hspace{0.3cm}}C\mbox{\hspace{0.3cm}}$ &1.0\mbox{\hspace{2.5cm}}
&0.5\mbox{\hspace{2.5cm}}&0.1\mbox{\hspace{2.5cm}}&0.05\mbox{\hspace{2.5cm}}\\
\hline
$\lambda$ & 0.96923323 & 0.99221794 & 0.99968755 & 0.99992188 \\
\hline
$\alpha$ & -0.26334376 & -0.04975108 & -0.00103697 & -0.00024860 \\
\hline
$\delta g_0 $ &-0.55711388 & -0.85478945 & -0.99357489 & -0.99831575 \\
\hline
$\delta g_1$ & 0.09131282 & 0.01196515 & 0.00000560 & -0.00000357 \\
\hline
$\delta g_2$ & 0.05735793 & 0.00736996 & -0.00001951 & -0.00000816 \\
\hline
$\delta g_3$ & 0.03704115 & 0.00553255 & 0.00003283 & 0.00000624 \\
\hline
$\delta g_4$ & 0.01757775 & 0.00255459 & -0.00000197 & -0.00000159 \\
\hline
$\delta g_5$ & 0.00679480 & 0.00111081 & 0.00000318 & 0.00000036 \\
\hline
$\delta g_6$ & -0.00025431 & -0.00000901 & -0.00000048 & -0.00000012 \\
\hline
$\delta g_7$ & -0.00374950 & -0.00063731 & -0.00000183 & -0.00000020 \\
\hline
$\delta g_8$ & -0.00472257 & -0.00087973 & -0.00000216 & -0.00000018 \\
\hline
$\delta g_9 $ & -0.00406673 & -0.00084036 & -0.00000206 & -0.00000016 \\ 
\end{tabular}
\vskip1cm
\caption{Dependence of the phase shift and the correction coefficients 
on the parameter $C$}
\label{tablazat}
\end{table}
\newpage
\begin{figure}

FIG. 1. The exact matrix elements $H_{j,k}$ in the intermittent case. \label{fig1a}

FIG. 2. The difference between the analytical approximation (\ref{e2.3}) 
and the exact values for the
 matrix elements $H_{j,k}$ in the intermittent case along the diagonal.
\label{fig1b}

FIG. 3. The exact matrix elements $H_{j,k}$ in the nearly intermittent case
($r=0.7$).\label{fig2a}

FIG. 4. Solid line: the difference between the analytical approximation (\ref{e3.2c}) 
and the exact values for the
 matrix elements $H_{j,k}$ in the nearly intermittent case $r=0.7$
along the maximum $k=\frac{2}{1+r}j$. Dashed line: the same 
for the difference between the analytical approximation (\ref{s3.3}) 
and the exact matrix elements.\label{fig2b}

FIG. 5. Solid line: the difference between the analytical approximation (\ref{e3.2c}) 
and the exact values for the
 matrix elements $H_{j,k}$ in the nearly intermittent case $r=0.7$
along the line $k=\frac{2}{3-r}j$. Dashed line: the same 
for the difference between the analytical approximation (\ref{s3.3}) 
and the exact matrix elements.\label{fig2c}

FIG. 6. Comparison of the analytical expression (\ref{s3.23}) with 
the results of a direct numerical measurement for the first
three nontrivial eigenvalues near $r=1$. Diamonds: $\lambda_1$, crosses:
$\lambda_2$, squares: $\lambda_3$ (numerically measured data). The dotted 
and solid lines display the corresponding 
analytical results up to first and second order in $\epsilon$, respectively.
\label{fig3}

FIG. 7. Comparison of the analytical expression (\ref{sfv1}) with 
the results of a direct numerical measurement for the first
nontrivial eigenfunction $s^{(1)}(x)$ at $r=0.99$. Solid line: analytical
approximation, dashed line: numerical result.
The eigenfunction is normalized such that the integral of its modulus is unity.
\label{fig4a}

FIG. 8. Same for the second eigenfunction $s^{(2)}(x)$.\label{fig4b}

FIG. 9. Same for the third eigenfunction $s^{(3)}(x)$.\label{fig4c}

FIG. 10. The ratio between the numerically determined correlation function 
and the analytical expression (\ref{kesz}) at three control parameter
values. Solid line: $r=0.9999$, dashed line: $r=0.99$, dotted line: $r=0.98$.
The functions $A(x)$ and
$B(x)$ are given by Eq.\ (\ref{c4}). \label{fig5}
\newpage\pagestyle{empty}
Fig.1. \epsfbox{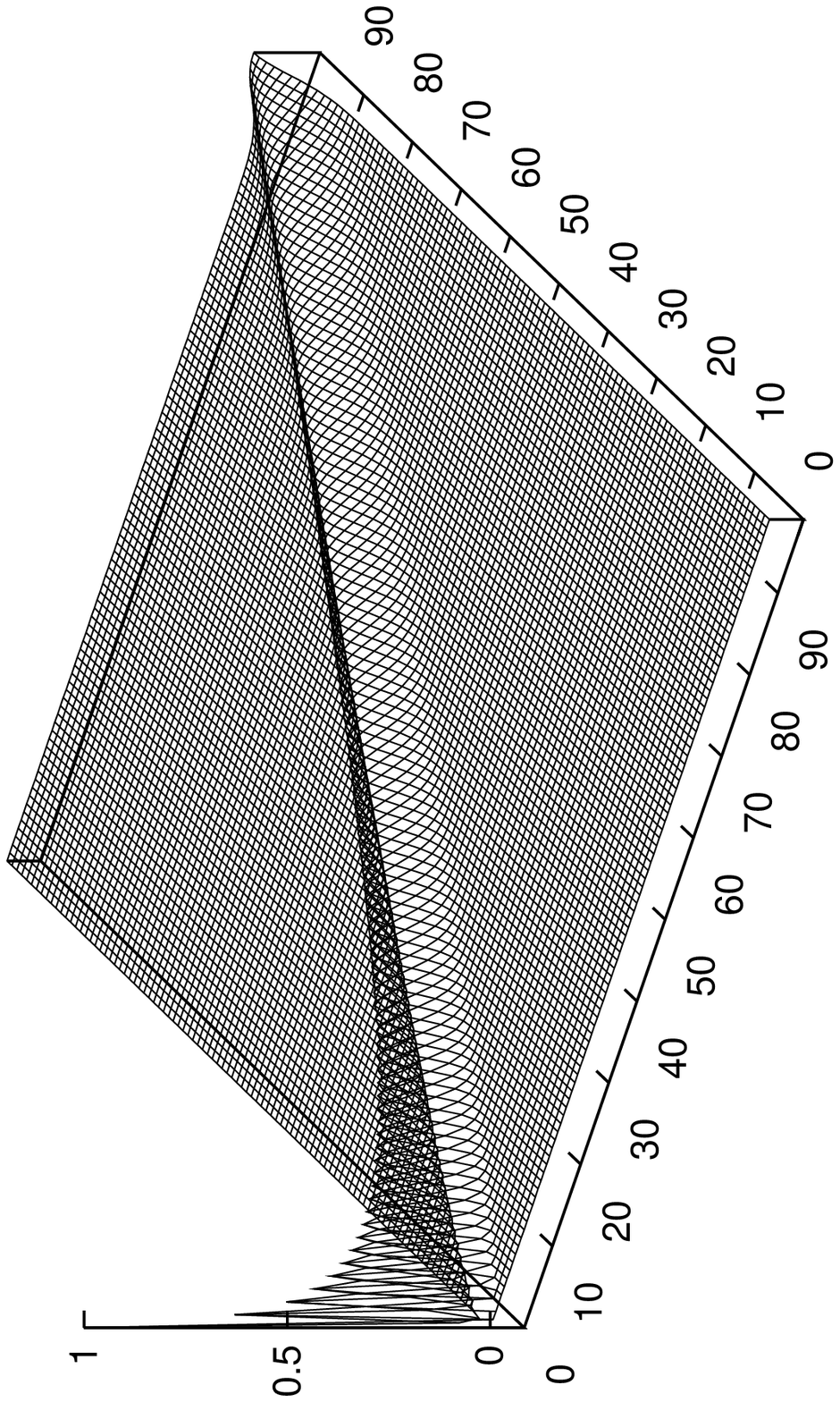}
\newpage \pagestyle{empty}
Fig.2. 
\epsfbox{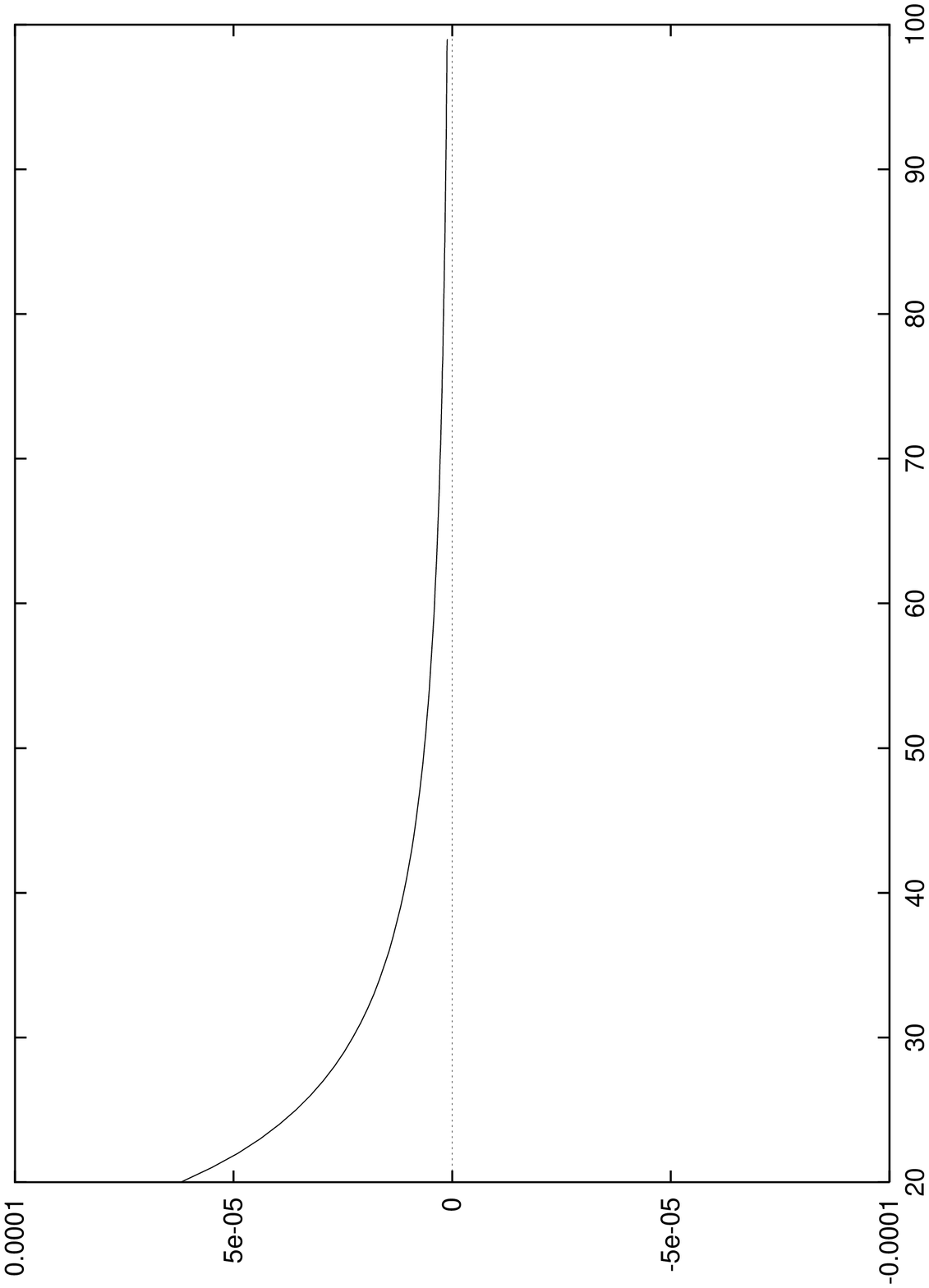}
\newpage\pagestyle{empty}
Fig.3. 
\epsfbox{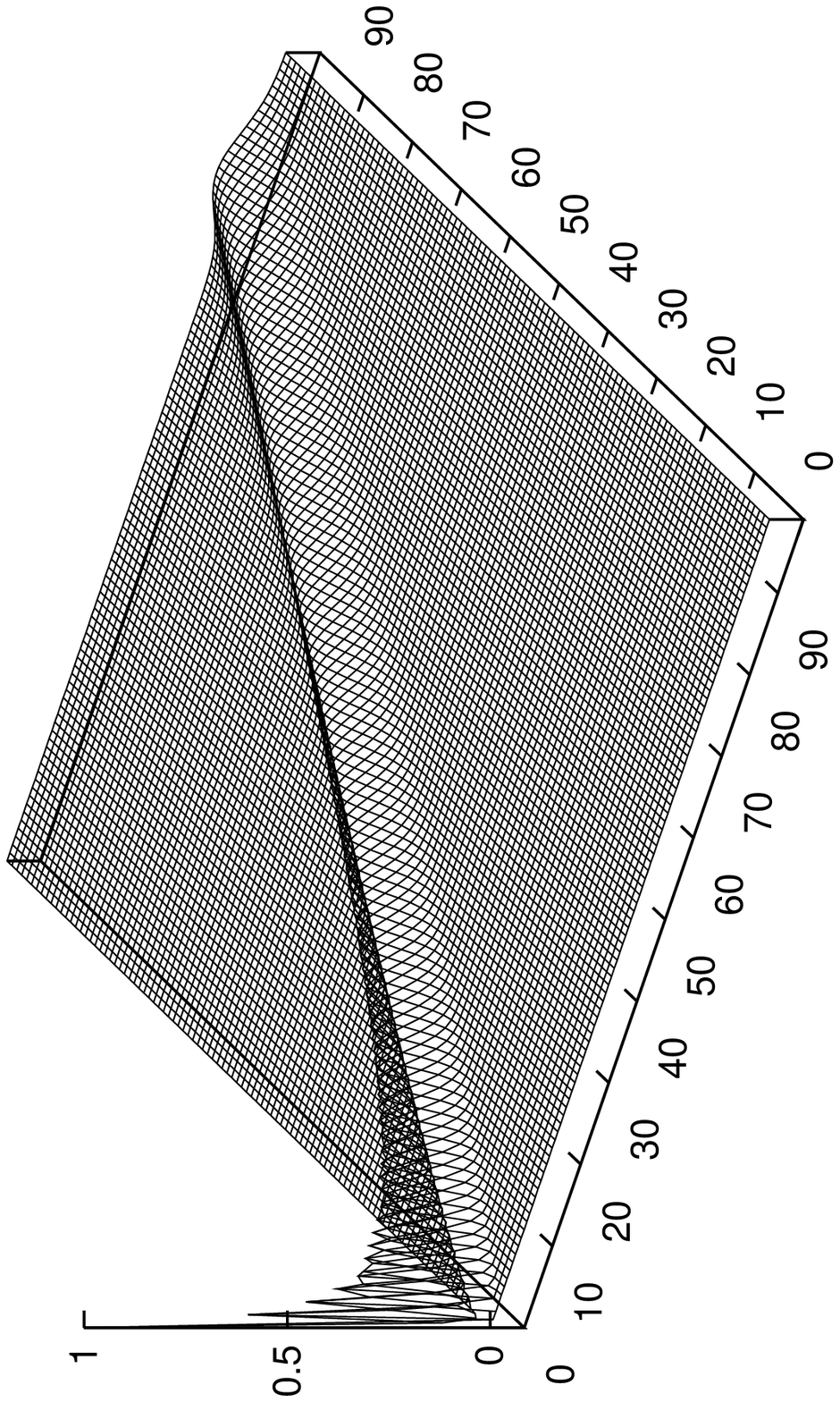}
\newpage\pagestyle{empty}
Fig.4.
\epsfbox{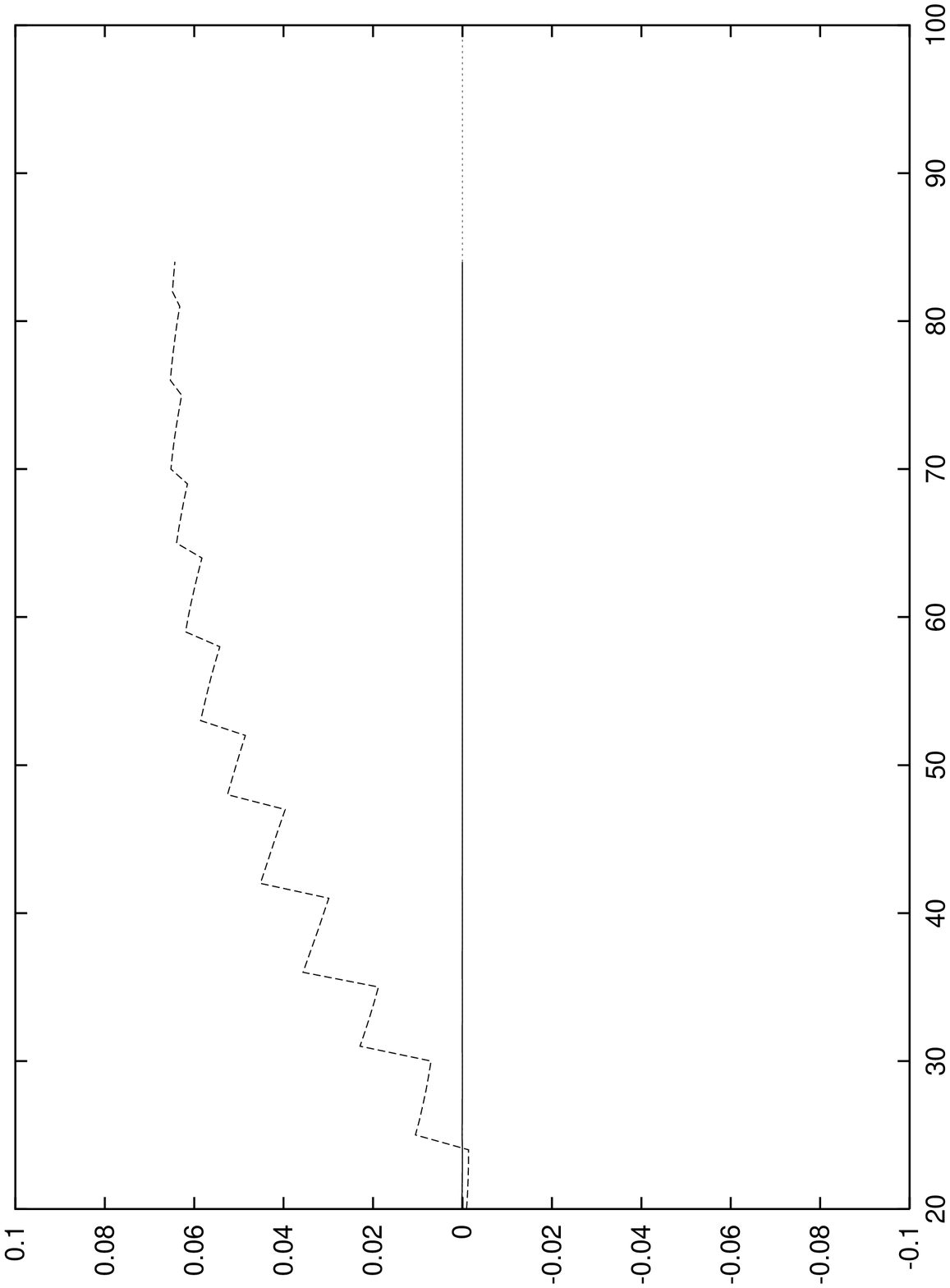}
\newpage\pagestyle{empty}
Fig.5.
\epsfbox{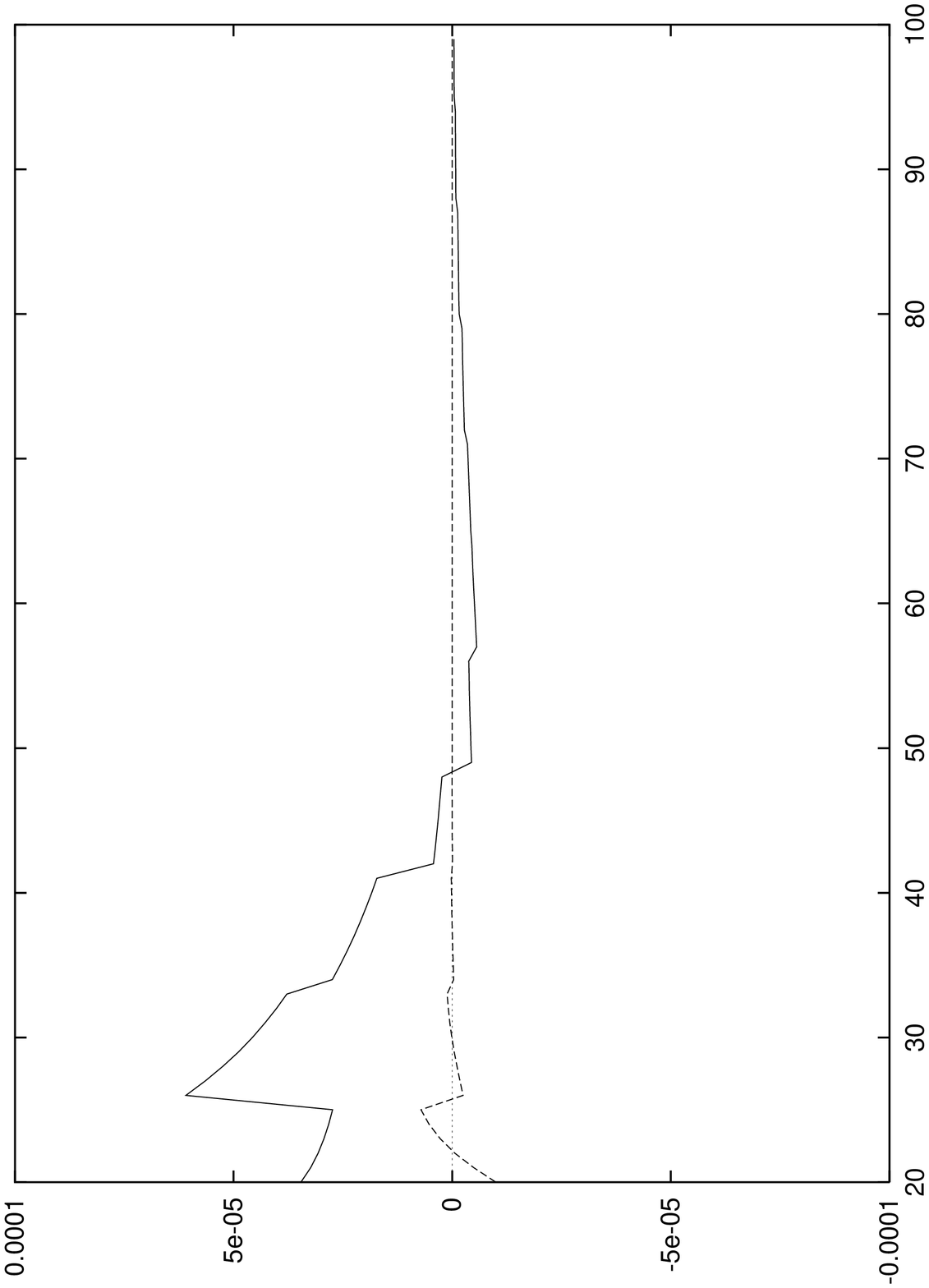}\newpage\pagestyle{empty}
Fig.6.
\epsfbox{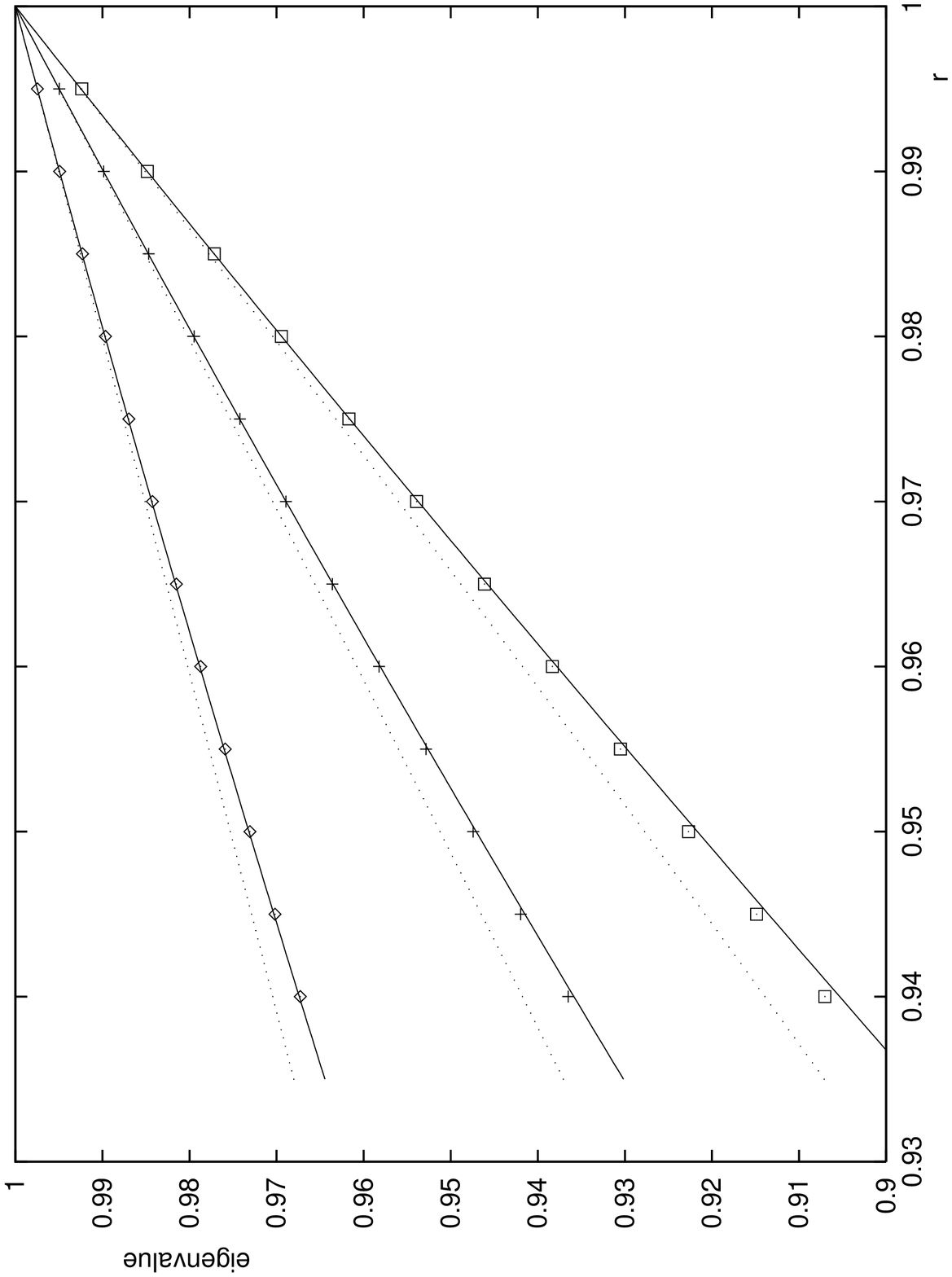}\newpage\pagestyle{empty}
Fig.7.
\epsfbox{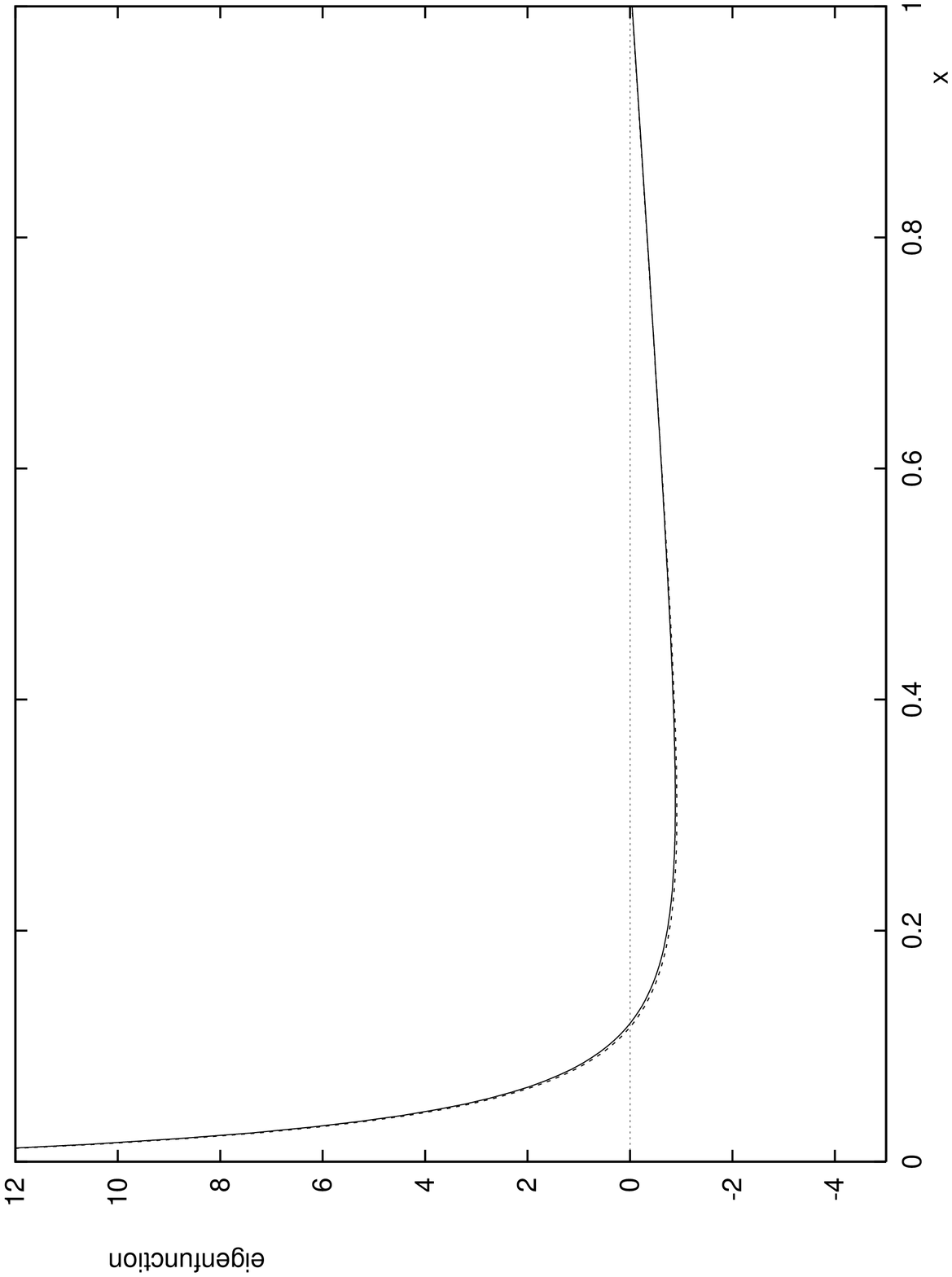}\newpage\pagestyle{empty}
Fig.8.
\epsfbox{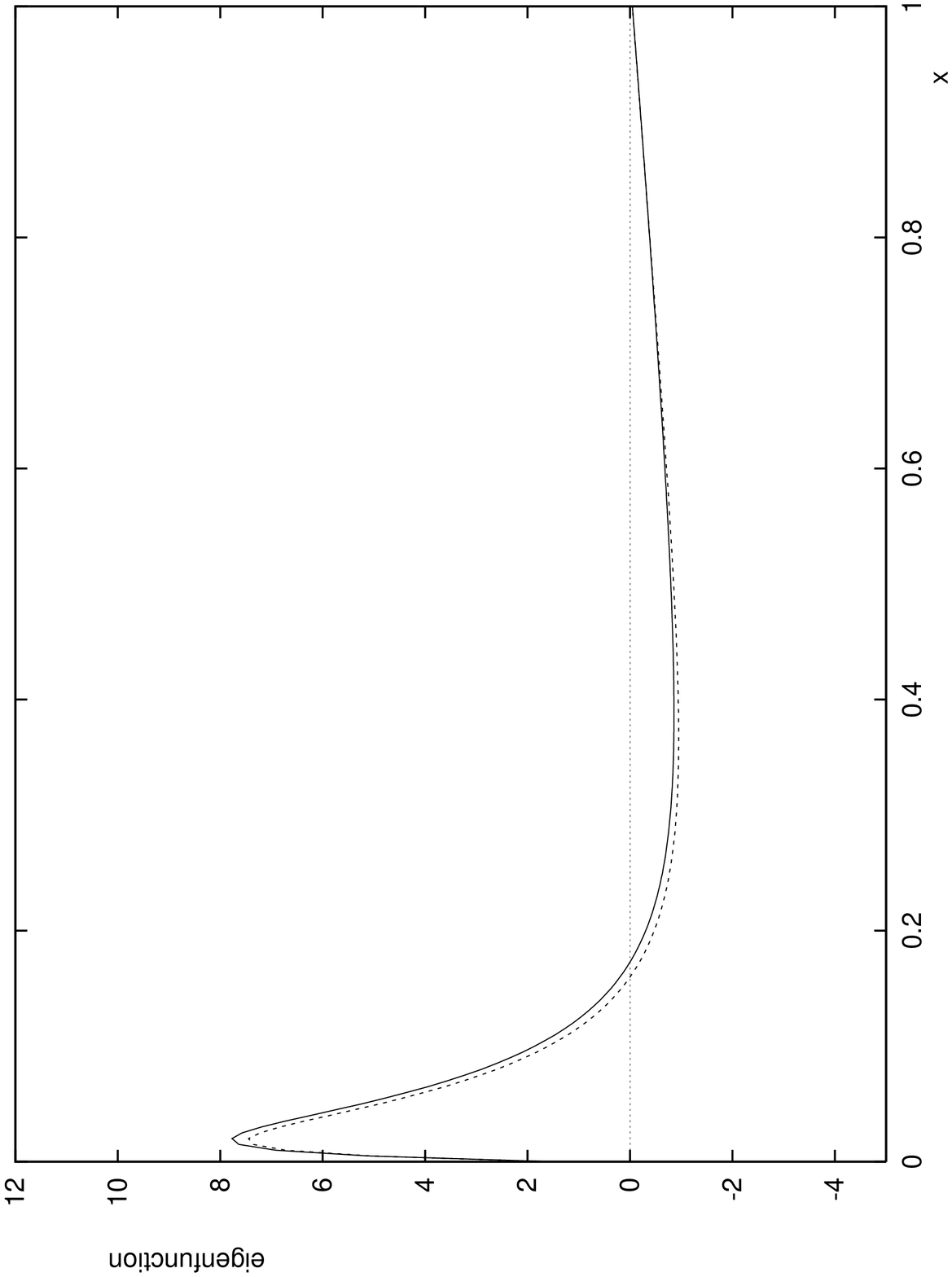}\newpage\pagestyle{empty}
Fig.9.
\epsfbox{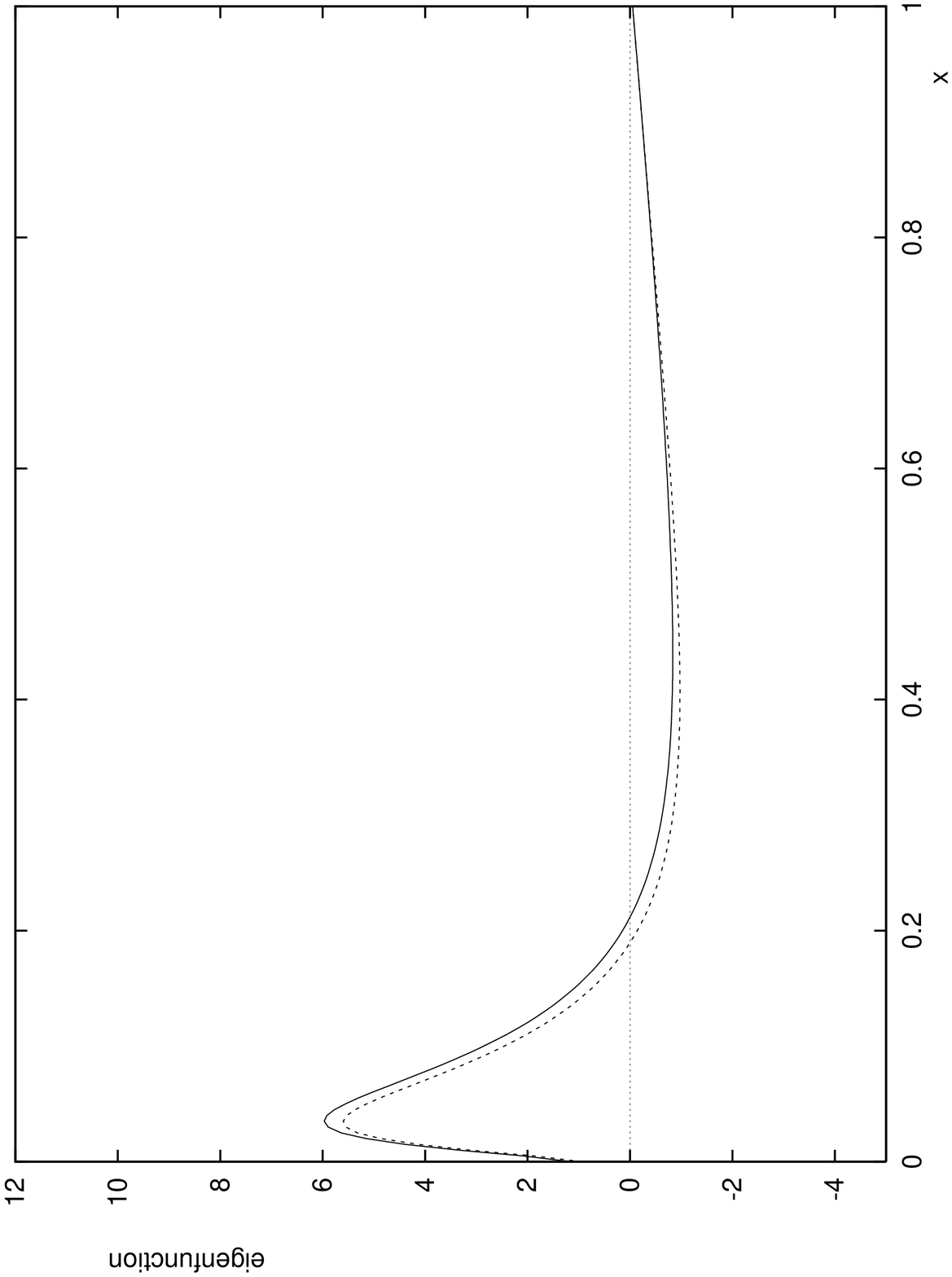}\newpage\pagestyle{empty}
Fig.10.
\epsfbox{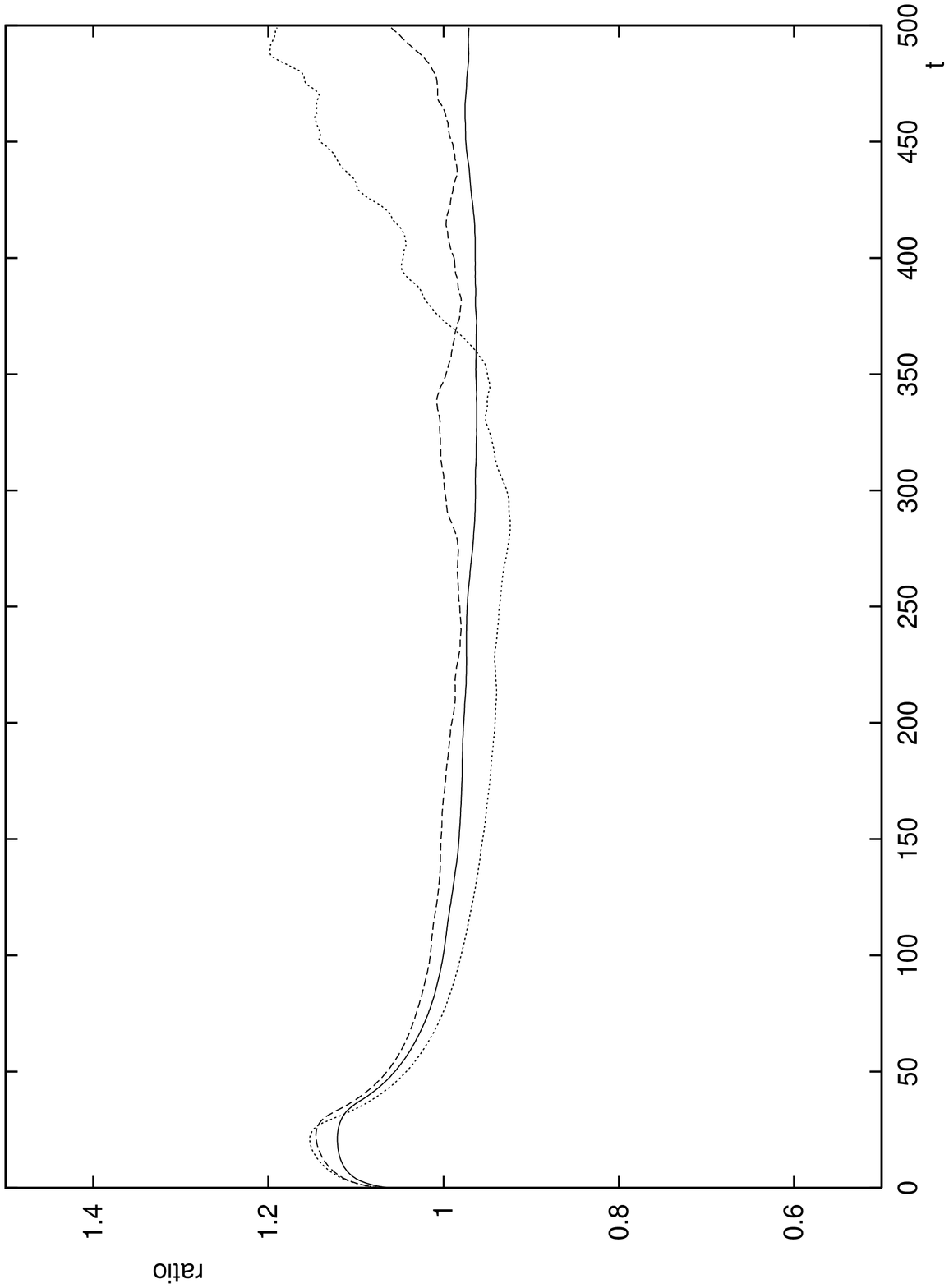}
\end{figure}
\end{document}